\documentclass[amsmath,amssymb,amsfonts,prc,showkeys,a4paper]{revtex4}
\usepackage{graphicx}
\usepackage{xcolor}

\begin{document}
\title{The critical behavior of hadronic matter:\\ Comparison of lattice and bootstrap model calculations\footnote{to appear in R. Hagedorn and J. Rafelski (Ed.), \emph{"Melting Hadrons, Boiling Quarks"}, Springer Verlag 2015}}

\author{L. Turko}
\email{turko@ift.uni.wroc.pl} \affiliation{Institute of Theoretical Physics,
University of Wroc{\l}aw, pl. Maksa Borna 9, 50-204  Wroc{\l}aw, Poland}

\begin{abstract}
{\hspace{3pt}Statistical bootstrap model and the related concept of the limiting temperature begun the discussion about phase transitions in the hadronic matter. This was also the origin of the quark-gluon plazma concept. We discuss here to which extend lattice studies of QCD critical behavior at non-zero chemical potential are compatible with the statistical bootstrap model calculations.}
\end{abstract}
\maketitle

\section{Rolf Hagedorn - some personal impressions}

"A fireball consists of fireballs, which in turn consist of  fireballs, and so on. . . ." - that was the leading sentence from the famous CERN Yellow Report 71-12 where Rolf Hagedorn presented in details leading ideas and results of his Statistical Bootstrap Model (SBM)\cite{haged:1971}. I met this Report in late 70' having yet some scientific experience both in quantum field theory as well as in the theory of high energy multiproduction processes.
 Starting from the beginning I've realized that I was reading something unusual. I was impressed by the elegance and precision of the presentation. It was quite obvious for me that the author had  spent a lot of time on discussions to clarify his arguments. Some questions were answered before I could even think about them. All was achieved without overusing of mathematical formalism, although all presentation was mathematically very rigorous. The author, however, used as simple and natural mathematical tools as possible, without going into complex jungle of formulae and multilevel definitions. It was also clear visible that the model, all its architecture and equipment is a one man project - Rolf Hagedorn.

And the most important point - the new idea was presented. I was not sure at that time - is this right or wrong one - but that was the idea not to be ignored. It was a nice answer for the long standing question - how effectively describe basic structure of matter, i.e. here hadronic matter. We knew the whole hierarchy - nuclei, nucleons, elementary particles, quarks … Any if those levels pretended at some time to be the "real" elementary one. The SBM didn't try to answer for the question about basic constituents. It just pointed out that this would be a wrong question.

I met Rolf Hagedorn in CERN in 1979.  It was about two years later. I was quite convinced yet that time to the idea of statistical bootstrap. I saw there also a good place to continue - at least as the way of thinking. I tried also to get some deeper knowledge of statistical physics which was earlier for me rather obscure subject in the domain of strong interactions. I also convinced my PhD student at that time, Krzysztof  Redlich, that this mixture of statistical physics and theory of elementary particles could be a very fruitful and interesting subject. Traveling to CERN I was quite excited to meet the physicist whose papers were giving me not only scientific but also quite esthetic experience.

In short: personal meetings with Hagedorn were even more interesting then reading his papers. He was a man of great general culture, very polite but also expecting well prepared arguments in discussions. From other side he was very open to present his reasoning, his calculations - even those being on a preliminary level. His hand-written notes were famous - in an almost calligraphic script, nicely written formulae, alternative arguments. He handed those notes to collaborators - it was as you received a chapter of an advanced textbook.

After two years our relations rapidly changed. The martial law, introduced in Poland in December 1981, not only made impossible my stay in CERN expected on Spring 1982, but also put me first in internee camp, then in jail. I was not only scientist at that time who found himself in such a unexpected surrounding. And it was Rolf Hagedorn, who without any delay, just in first days of  martial low, co-initiated in CERN campaign to free internee or jailed physicists in Poland. Posters with photos and names were posted on walls of TH division, signatures of protest were collected, letters of protest were sent to Polish officials.

When we met again in 1989 we still kept our relations, not only on scientific but also on a friendly level. Looking now back I must admit that Rolf Hagedorn was among those who shaped my profile - not only as a scientist, but also as a man. He was definitely worth to follow - in any respect.  I am very happy I had the possibility to be close with such an exceptional scientist and an exceptional man. Man of honor.

\section{Critical Behavior of Hadronic Matter}
Quantum chromodynamics (QCD) gauge theory is an excellent tool for description of
single hadronic events in vacuum. However, for the dense and hot hadronic matter the
most reliable theoretical results, based on first principles, can be obtained only
through lattice gauge theory calculations. In particular, phase transitions or
crossover phenomena are expected. These critical behavior is related to
peculiarities in standard lattice QCD quantities as the Polyakov loop or
susceptibilities.

From other side, surprisingly simple resonance gas model provides a good description of particle yields in the
relativistic heavy ion collisions in the broad energy range \cite{cleym_redl:1998,braun_mag:2001}. The clue to this result is in the exponential-like behavior of the particle mass spectrum. This model, slightly extended \cite{Blaschke:2015nma}, reproduces also results of the transition between a hadron resonance gas phase and the quark gluon plasma obtained in course of QCD lattice simulation.

The concept of the limiting temperature of the hadronic matter has appeared  in the
statistical bootstrap model (SBM) \cite{haged:1965, hagraf:1981, frautschi, haged:1994}. An introduction of the baryonic
chemical potential transforms this critical temperature into the critical curve.
Higher internal symmetries lead to an appearance of the critical surface
\cite{red-tur:1980, Turko:1981nr}. The hadronic matter, above the critical curve, is
interpreted nowadays as a quark-gluon plasma (QGP) phase.

I will compare here calculations of critical curves obtained in the $\mu-T$ plane
from lattice Monte Carlo simulations with analogous critical curves obtained with the same
input from the statistical bootstrap model. It was shown
\cite{kars_redl:2003,kars:2004} that a gas of non-interacting resonances provides a
good description of the low temperature phase of lattice QCD. As the hadronic mass
spectrum is similar to the exponential mass spectrum expected from the SBM, it is
interesting to check if the critical behavior obtained from the SBM resembles Monte
Carlo results of lattice QCD.

Hot and dense hadronic matter undergoes a transition to a deconfined and chirally symmetric medium at given combinations of temperature and baryon chemical potential.

We are interested here in the region of the $(\mu_B, T)$ plane covered by ultrarelativistic heavy ion collisions where the phase transition is expected i.e.  the high temperatures and low baryonic densities. The efficient method of lattice simulation is here via a Taylor expansion with respect to the baryonic chemical potential at $\mu_B = 0$ \cite{Gottlieb:1988ga, Gavai:2002gg}. This lattice technique, supplemented with other technical tools specific for QCD lattice simulations, was used to obtain the phase transition curve $T_c(\mu)$ for 2-flavor and 3-flavor QCD \cite{Allton:2002zi,Karsch:2001nf,Karsch:2000kv, Karsch:2003va,Redlich:2004gp}.

\subsection*{Critical curve from the lattice calculations}

In order to compare the SBM with lattice results one should take into
account that the latter are not obtained from calculations performed with the
physically realized quark mass spectrum. One finds \cite{rich,gock,Zanotti:2003fx} that the quark
mass dependence is well parameterized through the relation
\begin{equation}\label{quarkmass}
 (m_H a)^2 = (m_H a)^2_{phys} + b(m_\pi a)^2\,,
\end{equation}
where $(m_H a)_{phys}$ denotes the physical mass value
of a hadron expressed in lattice units and $(m_H a)$ is the value calculated on the
lattice for a certain value of the quark mass or equivalently a certain value of the
pion mass.

  Lattice constant $a$ can be treated as a specific ultra-violet regularization which is removed in the continuous limit $a\to 0$. The value of the critical temperature $T_c$  is dependent on the pion mass \cite{Karsch:2000kv}. Pion here is understood as the lowest pseudoscalar mesonic state $q\bar q$ of the mass $m_{PS}$. This mass decreases to its physical pion mass $m_\pi = 0.140$~GeV in the continuous limit along with the critical temperature.

  \begin{figure}[h!]
\centerline{
\includegraphics[width=0.42\textwidth,angle=0]{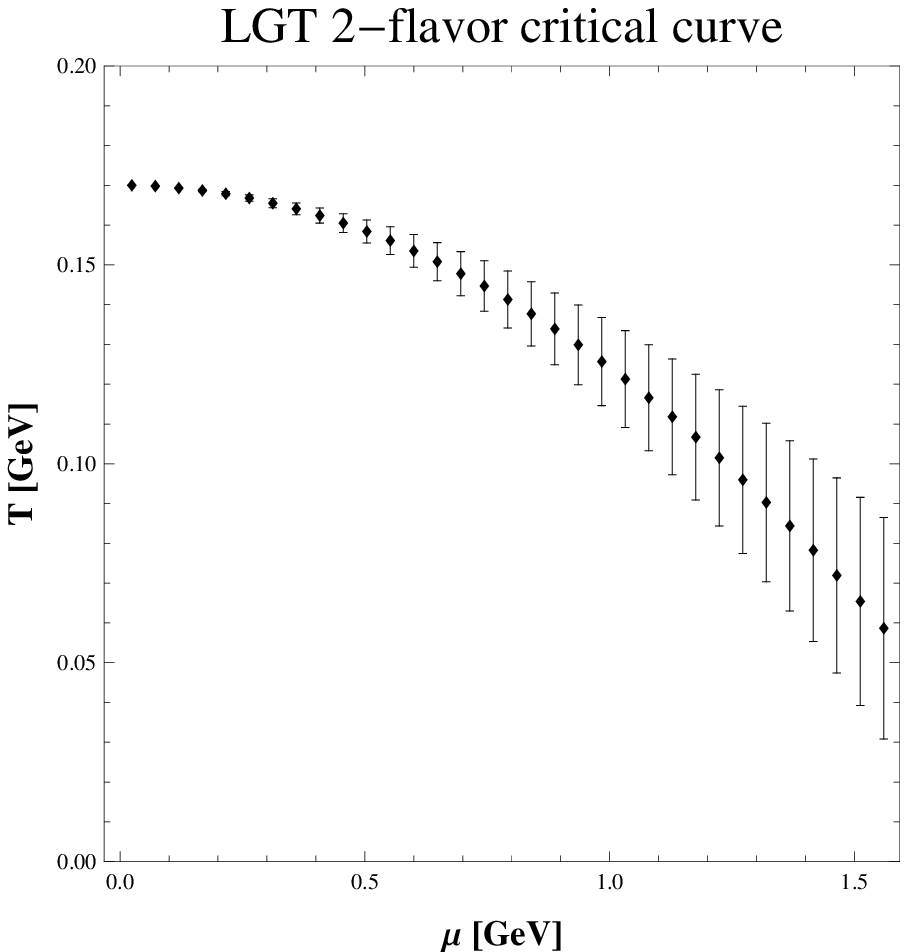}\hfill\includegraphics[width=0.4\textwidth,angle=0]{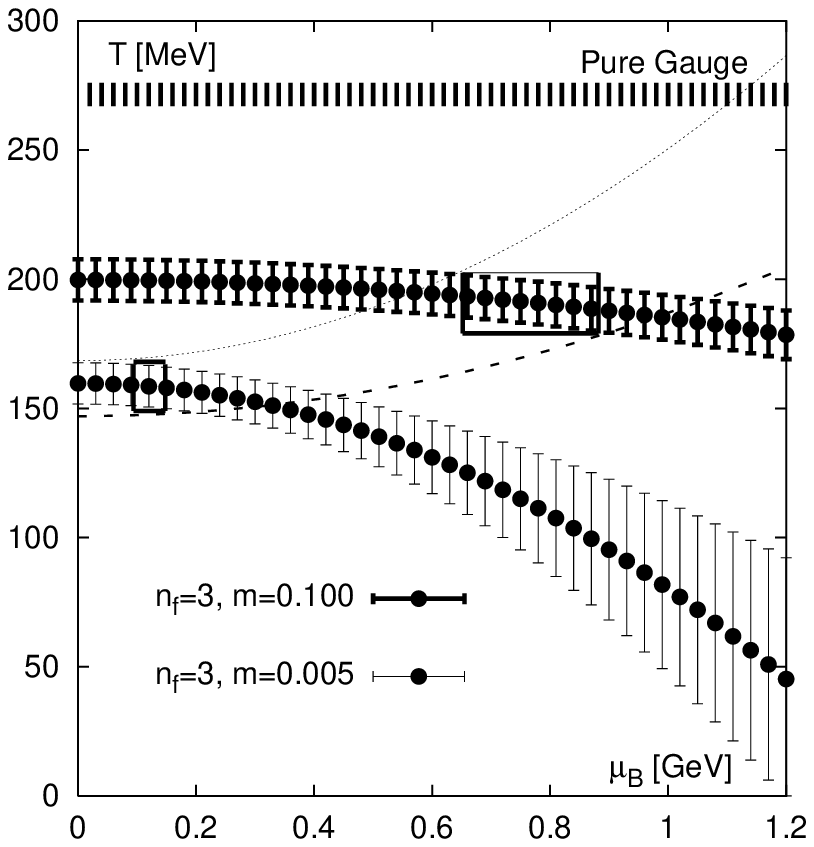}}
\caption{\label{Tc2}The transition temperature $T_c$ , as a function of baryonic chemical potential for 2-flavor (left-hand figure) and 3-flavor QCD (right-hand figure) lattice simulation.
}
\end{figure}

Critical curves from Fig.\ref{Tc2}  were obtained at some assumed quarks masses (in lattice constant $a$ units):  $m_q=0.1$ on the left-hand figure and $m_q=0.1,\ m_q=0.005$ on the right-hand plots. Corresponding $m_{PS}$ masses were $0.770$~GeV, $0.190$~GeV and $0.170$~GeV respectively.
\subsection*{Critical curve from the statistical bootstrap model}
Let us start from the bootstrap equation taken for the system with pions and
nucleons taken as basic constituents. The bootstrap input function is given as
\begin{equation}\label{bootstr_inp1}
    \varphi_{n_\pi,n_N}(\mu,T) = 2 H \pi T \left[n_\pi m_\pi K_1\left(\frac{m_\pi}{T}\right) + n_N m_N
K_1\left(\frac{m_N}{T}\right)\cosh\left(\frac{\mu_B}{T}\right)\right]\,.
  \end{equation}

I restricted the set of input particles for a given valence quark input to lightest
mesonic and baryonic states respectively. $n_\pi$ and $n_N$ are numbers of lightest mesonic  and baryonic states respectively for a given valence quark input. Spin degeneration and antibaryons are taken into
account here. They form an input for the SBM. So for two quark flavors there are
$n_\pi=3$ mesonic states and $n_N=8$ baryonic states. For three quark flavors with
the threefold quark mass degeneracy one gets $n_\pi=8$ and $n_N=32$, respectively.

The bootstrap constant $H$ is written as

\begin{equation}\label{bstrH}
  H = A\frac{2m_\pi m_N}{(2\pi)^3}\frac{1}{B}
\end{equation}

where $B^{1/4}\approx 0.190$ MeV is the bag constant to reproduce critical energy density
$\varepsilon\approx 0.6$~GeV/fm$^3$ and the parameter $A$ is chosen so to get the critical temperature $T_c$ at $\mu_B = 0$ from the corresponding QCD lattice simulation.

The statistical bootstrap model used on the QCD lattice system has its basis components such as they appear in lattice QCD simulation. It means, particularly, nucleon mass expressed by pion mass (all in GeV) as
\begin{equation}\label{mn:mass}
  m_N(m_\pi) = 0.94  + \frac{m_\pi^2}{0.94}
\end{equation}

The critical curve is obtained directly from the bootstrap equation
\begin{equation}\label{sbm equ}
    2\Phi=\varphi+e^\Phi-1\,,
\end{equation}
which is meaningful only for
\[\varphi\leq\ln 4-1\,.\]
So the critical curve $T_c(\mu)$ is given on the $\mu_B-T$ plane by the condition
\begin{equation}\label{sbm crit}
    \varphi_{n_\pi,n_N}(\mu,T)=\ln 4 -1\,.
\end{equation}

\subsection*{Comparison of SBM and lattice-QCD}

\begin{figure}[h!]
\centerline{
\includegraphics[width=0.4\textwidth,angle=0]{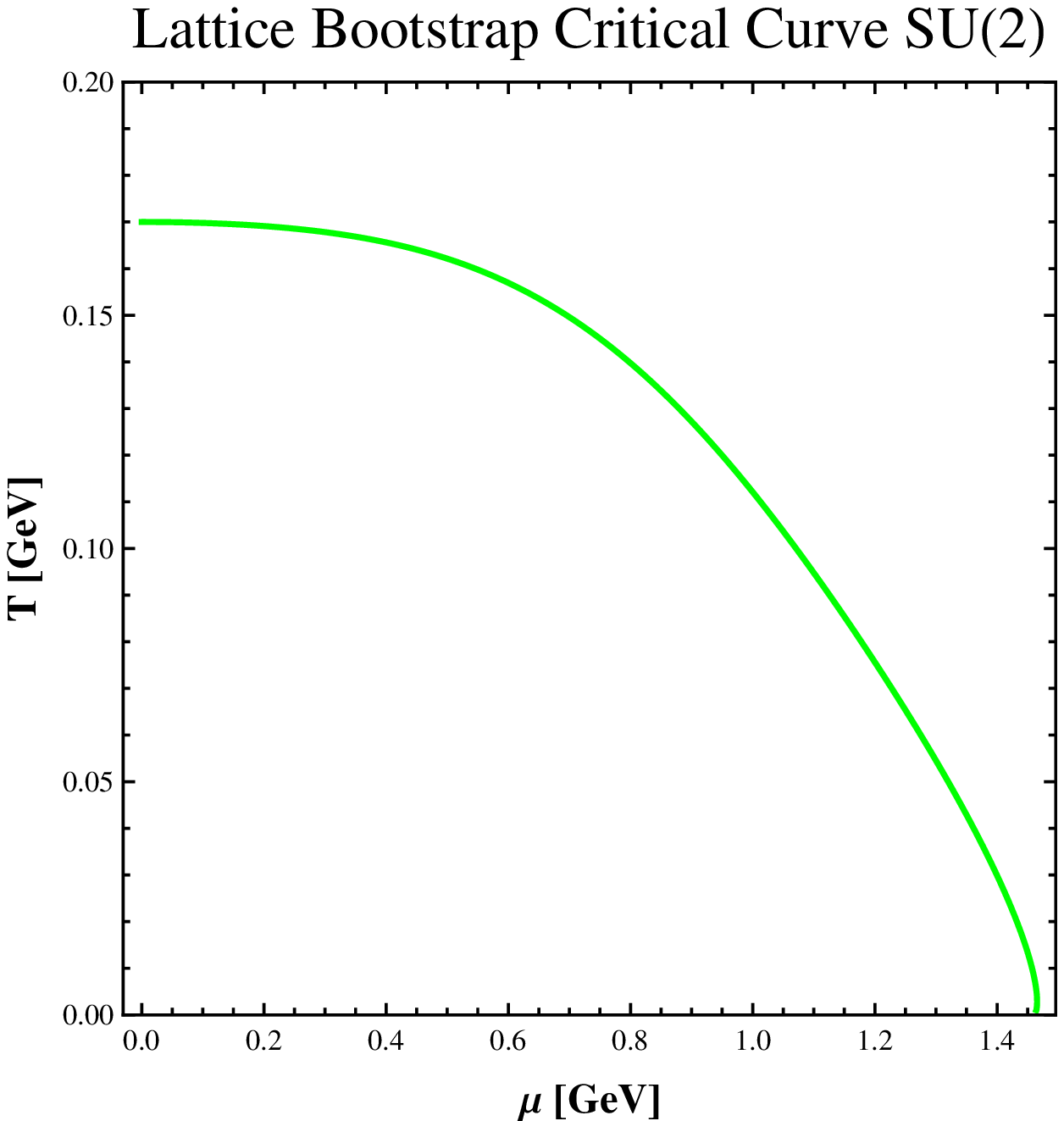}\hfill\includegraphics[width=0.4\textwidth,angle=0]{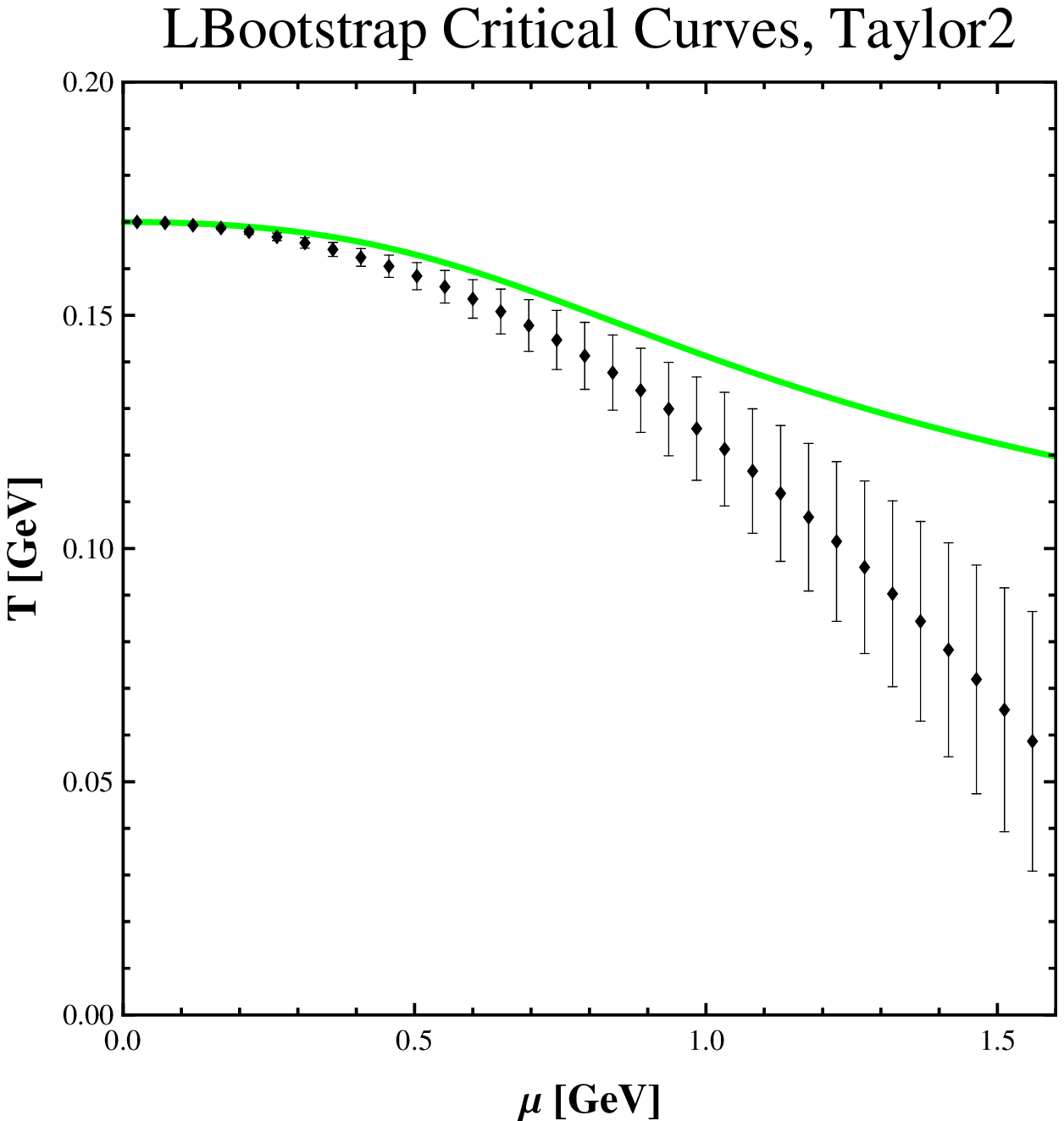}}
\caption{\label{LTc2}The transition temperature $T_c$ , as a function of baryonic chemical potential for 2-flavor lattice, bootstrap system (left-hand figure). The critical temperature 2-flavor lattice bootstrap system, compared with result of corresponding lattice simulation (right-hand figure), both considered in power law approximation.
}
\end{figure}

Let us consider lattice bootstrap system corresponding to the 2-flavor QCD lattice simulation from the Fig.\ref{Tc2}. The critical curve resulting from the statistical bootstrap model is depicted on the Fig.\ref{LTc2} - left panel.

As the critical curves $T_c(\mu_q)$ from the QCD lattice calculations were obtained
up to $\mathcal{O}((\mu_q/T_c(0))^2)$ term, so the similar approximation should be
used for the critical curves obtained from Eq. \eqref{sbm crit}. This means that the expression
$\cosh[\left(\frac{\mu_B}{T}\right)]$ in Eq. \eqref{bootstr_inp1} should be replaced by the corresponding Taylor expansion truncated to the first two terms.

The result of this procedure is presented on the Fig.\ref{LTc2} - right panel.

\section{Conclusions}
Results presented on Fig.\ref{LTc2} show that statistical bootstrap model reproduces at least qualitatively basic properties of the critical curve obtained in the course of QCD lattice simulation. We have quantitative agreement for smaller values of baryonic chemical potential, not exceeding $0.7$ GeV. This is rather natural taking into account the method used in the simulations, based on the idea of analytical continuation in the chemical potential variable, starting from the point $\mu=0$.

The statistical bootstrap model, created by Rolf Hagedorn half of century ago, in the time when quarks were still a bold hypothesis, still remains a very inspiring research tool of hadronic matter. Based on the deep knowledge and great Hagedorn's intuition the model has still some unknown and unexpected properties, waiting to be discovered.

\begin{acknowledgments}
I acknowledge the stimulating discussions with F.~Karsch, J.~Rafelski, and K.~Redlich. This work has been supported by the Polish National Science Center under grant no. DEC-2013/10/A/ST2/00106.
\end{acknowledgments}

\end{document}